\documentclass[letterpaper]{JHEP}

\usepackage{amssymb,amsmath}
\usepackage{epsfig}
\usepackage{epstopdf}

\def\be{\begin{eqnarray}}       
\def\ee{\end{eqnarray}}

\def\3{\ss}

\def\half {\frac{1}{2}}

\preprint{{\tt hep-th/0607037}}

\title{Toy models for closed string tachyon solitons}

\author{Oren Bergman and Shlomo S.~Razamat\\
Department of Physics\\
Technion, Israel Institute of Technology\\
Haifa 32000, Israel\\
\email{bergman, razamat@physics.technion.ac.il}}

\abstract{We present qualitative evidence that closed string tachyon solitons describe
backgrounds of lower-dimensional sub-critical string theory.
We show that a co-dimension one soliton in the low energy effective
gravity-dilaton-tachyon theory in general has a flat string-frame metric, and
a dilaton that grows in both directions away from the core, and is linear
in the soliton worldvolume coordinates.
Spacetime, as seen in the Einstein frame, is therefore effectively localized
in $(D-1)$-dimensions, in which the dilaton is linear, in agreement with
the linear dilaton background of the $(D-1)$-dimensional sub-critical string.
We construct a number of exactly solvable toy models with specific tachyon
potentials that exhibit these features,
and address the question of finding solitons in the bosonic closed string field theory
using the recent advances in computing the tachyon potential.}

\keywords{Tachyon condensation}

\begin{document}


\section{Introduction}

The role of the closed string tachyon in bosonic string theory 
(or in Type 0 superstrings) is one of the oldest unsolved puzzles 
in string theory. Although much progress has been made
in understanding the dynamics of open string tachyons 
\cite{Sen:2004nf,Taylor:2003gn},
and localized closed string tachyons 
\cite{Adams:2001sv,Vafa:2001ra,Harvey:2001wm,Okawa:2004rh, Headrick:2004hz,Bergman:2004st,Adams:2005rb,
Horowitz:2005vp,Bergman:2005qf},
bulk closed string tachyons have remained a mystery.

Sen's two-part conjecture was the guiding principle in the study of open string tachyons
in unstable D-branes.
The first part of the conjecture told us that the ground state is the closed string 
vacuum, 
and the second part told us that (stable) solitons are lower-dimensional (stable) 
D-branes. The first part is usually sub-divided into the conjecture 
identifying the height of the tachyon potential
with the tension of the D-brane, and the conjecture that there are no open string excitations
in the ground state.
These conjectures have been subjected to extensive quantitative tests,
and a proof of the first part has recently been presented in \cite{Schnabl:2005gv,Ellwood:2006ba}.

There is a very similar picture for localized closed string tachyons.
Tachyonic instabilities living on or near space-time singularities, 
like orbifolds, black holes or linear dilatons,
have the effect of smoothing out the space by removing the part of spacetime containing
the singularity,
and with it the closed string states localized near the singularity.
Though we have not yet understood the role of solitons of localized closed string
tachyons, it is likely that they correspond to lower dimensional singularities.

\medskip

This leads to 
a natural two-part conjecture for bulk closed string tachyons:
\begin{enumerate}
\item The ground state contains no degrees of freedom (perhaps a topological theory).
\item Solitons correspond to lower-dimensional sub-critical closed string theories.\footnote{Indirect evidence
for this conjecture was given in the context of p-Adic strings by Moeller and Schnabl \cite{Moeller:2003gg}.}
\end{enumerate}

The convential techniques used to analyze tachyon condensation in the open 
string case, and to some extent in the localized closed string case,
are much less effective when applied to the case of bulk closed string tachyons. 
For example worldsheet RG analysis, which was very successful in verifying
open string tachyon conjectures \cite{Harvey:2000na, Gerasimov:2000zp,Kutasov:2000qp}, 
as well as localized closed string tachyon conjectures
\cite{Adams:2001sv,Vafa:2001ra,Harvey:2001wm,Headrick:2004hz}, 
is difficult to apply to bulk closed string tachyons.\footnote{See however
\cite{Hellerman:2004zm,Hellerman:2004qa} for a worlsheet approach 
to closed string tachyon condensation in super-critical heterotic string theories,
which is used to argue that they decay to the critical heterotic theory.}
The central charge must decrease
along the RG trajectory, and this contradicts the $c=0$ condition of string theory.
On the other hand, string field theory, which has also been extremely useful in the case
of open string tachyons is much more difficult to apply to closed string tachyons. 
This is bacause closed string
field theory (CSFT) is non-polynomial; it contains an infinite number of interaction
vertices \cite{Zwiebach:1992ie}. Nevertheless CSFT has been used to successfully
verify, at a qualitative level, the conjectures associated with some of the localized
closed string tachyons \cite{Okawa:2004rh,Bergman:2004st}. This gives some hope that it might also 
shed light on the bulk closed string tachyon.
A major step in this direction has been taken by Yang and Zwiebach, who computed 
corrections to the closed string tachyon potential due to higher level 
fields \cite{Yang:2005rx}. Their result supports the existence of a non-trivial critical point.
 
\medskip

In this paper we will investigate part 2 of the conjecture for the bulk closed string
tachyon of the bosonic string. In particular, we would like to find a co-dimension one
soliton solution of the 26-dimensional closed bosonic string theory, and compare it with
the 25-dimensional sub-critical bosonic string theory.

In section 2 we will first formulate the problem in the low-energy effective theory of 
the tachyon, dilaton and graviton.
To specify the action we have to know the form of the tachyon potential. 
However some general features of the soliton solution will be independent of
the specific potential. 
We will show that the string metric is flat, and that the dilaton is linear in the 
coordinates along the soliton, and grows in the transverse coordinate as we move away from 
the core of the soliton.
Spacetime as measured in the Einstein frame therefore collapses away from the soliton,
as it did in the time dependent tachyon background of \cite{Yang:2005rw}.
We will also show that the lowest mode of the dilaton in this background
decouples from the tachyon, and has a mass given by the slope of the dilaton
along the soliton.
All this is consistent with the identification of the soliton as the linear dilaton
vacuum of the 25-dimensional sub-critical string theory.

In section 3
we will present three exactly solvable tachyon-dilaton toy models as examples of these general features.
In each of these models we will "derive" a tachyon potential starting from a simple ansatz
for either the tachyon or dilaton profile of the soliton, and then solve for the other.

In section 4  we begin an investigation of soliton backgrounds in CSFT,
using the recently obtained results for the potential in \cite{Yang:2005rx}.
Our results here are preliminary, and there is definitely room for improvement.
However the indication is that the tachyon lump of the lowest-level cubic
potential survives when we include higher level fields and higher point vertices.

\section{Generic features of a closed string tachyon soliton}

The low energy effective theory of the gravity-dilaton-tachyon system
in the closed bosonic string (or the NSNS sector of the Type 0 superstring)
is given by
\be\label{action}
S=\frac{1}{2\kappa^2}\int d^{D}x\sqrt{-g}\,
 e^{-2\Phi}\left(R+4(\partial_\mu\Phi)^2-(\partial_\mu T)^2-2V(T)\right)\, ,
\ee
where $V(T)={1\over 2}m^2 T^2 + \cdots$.
The equations of motion for the metric, tachyon and dilaton are given respectively by
\be
R_{\mu\nu}+2\nabla_\mu\nabla_\nu\Phi-\partial_\mu T\partial_\nu T&=&0 \\
\nabla^2 T-2\partial_\mu \Phi\partial^\mu T-V'(T) &=& 0 \\
\nabla^2 \Phi-2\partial_\mu \Phi\partial^\mu\Phi -V(T) &=& 0 \,.
\ee

\subsection{Co-dimension one soliton}

We would like to find a co-dimension one soliton, namely a static solution 
$\bar{T}(x_1)$,
such that $\bar{T}(0)=0$ and $\bar{T}(x_1)$ approaches a minimum of $V(T)$ as 
$x_1\rightarrow \pm\infty$. If there is a unique minimum, the soliton is a lump,
and if there are degenerate minima the soliton is a kink. We will assume that the
metric has the following form
\be
ds^2=dx_1^2+a(x_1)^2\eta_{\mu\nu}dx^\mu dx^\nu \,,
\ee
where $\mu,\nu = 0,2,\ldots,D-1$.
The dilaton will depend on all the coordinates in general, but using
the $(D-1)$-dimensional Lorentz symmetry we can fix $\Phi = \bar{\Phi}(x_1,x_2)$.
The different components of the gravity equation then reduce to
\be
\mbox{} -(D-1)a^{-1}a'' + 2\partial_1^2\bar{\Phi} - (\bar{T}')^2 &=& 0 \label{11} \\
\mbox{} - a a'' - (D-2)(a')^2
  +2 a a' \partial_1\bar{\Phi} &=& 0\label{mn}\\
\mbox{} - a a'' - (D-2)(a')^2 +2\partial_2^2\bar{\Phi}
 +2a a' \partial_1\bar{\Phi} &=& 0\label{22}\\
\partial_1\partial_2\bar{\Phi}-a^{-1} a' \partial_2\bar{\Phi} &=& 0 \,.\label{12}
\ee
From (\ref{mn}) and (\ref{22}) we see that $\partial_2^2\bar{\Phi}=0$,
and from (\ref{mn}) we see that $\partial_1\bar{\Phi}$ is independent of $x_2$,
so we deduce that
\be
\label{total_dilaton}
 \bar{\Phi}(x_1,x_2) = D(x_1) + Qx_2 \,,
\ee
for some $Q$. 
The dilaton must therefore be linear in the soliton "worldvolume" coordinates.
Equation (\ref{12}) then gives
\be
 Q a' = 0 \,,
\ee
so either $Q=0$, or $a(x_1)$ is a constant which we can set to 1,
and the string metric is flat.
We will assume the latter.
In this case the remaining gravity equation (\ref{11}), and the
tachyon and dilaton equations reduce respectively to
\be
 2 D'' - (\bar{T}')^2 &=& 0 \label{gravity}\\
\bar{T}'' - 2 D' \bar{T}'-V'(\bar{T}) &=& 0 \label{tachyon}\\
D'' - 2(D')^2 - 2Q^2-V(\bar{T}) &=& 0 \label{dilaton} \,.
\ee
This is an over-determined set of equations for the two fields.
In particular we get two independent expressions for $D'$,
\be
D' = \half\frac{\bar{T}''-V'(\bar{T})}{\bar{T}'} 
  =\half\sqrt{(\bar{T}')^2-4Q^2-2V(\bar{T})} \,.
\ee
This fixes the value of $Q$ in terms of the parameters of the tachyon potential.

We saw that the dilaton will generically be linear along the soliton.
The qualitative behavior of the dilaton in the transverse coordinate $x_1$
is also generic. Consider the tachyon equation (\ref{tachyon}). 
Regarding $x_1$ as time, this is the equation of motion of a particle 
with position $\bar{T}$ moving in a potential $-V(\bar{T})$ with friction $-2 D'$.
In this language the soliton is described by the motion where the particle 
begins at $x_1\rightarrow -\infty$ at a maximum of $-V$, rolls 
down and up (and then back down and up if the maximum is unique),
and ends at a maximum at $x_1\rightarrow \infty$. 
For this to be a solution
$D'$ cannot be monotonic. There has to be as much negative friction 
as positive friction in order to conserve the total energy.
This, together with the gravity equation (\ref{gravity}) which implies that
$D'' \geq 0$, and the requirement of smoothness,  shows
that the dilaton must grow in both directions away from the soliton.
Since the string metric is flat, this implies that spacetime, measured in
the Einstein frame, collapses away from the soliton.
It is effectively localized on the $(D-1)$-dimensional 
worldvolume of the soliton.
The whole picture is consistent with the identification of the soliton
as the flat linear-dilaton background of the $(D-1)$-dimensional string theory.

\subsection{Fluctuations}

The analysis of the fluctuations in this system is very complicated.
Even if we ignore the metric fluctuations and consider only the dilaton
and tachyon fluctuations
\be
\delta T = T - \bar{T} \;,\; \delta\Phi = \Phi - \bar{\Phi} \,,
\ee
the mixing terms in the quadratic action are non-trivial:
 \be
 \frac{1}{2\kappa^2}\int d^{D}x\, e^{-2\bar{\Phi}}&&
 \bigg\{ - 4 V(\bar{T}) + 4(\partial_\mu \delta\Phi)^2-(\partial_\mu \delta T)^2
 -V''(\bar{T})\delta T^2 + 4V'(\bar{T})\delta T \delta\Phi \nonumber\\
 && \mbox{} \left[ 8(D')^2 + 8Q^2 - 2(\bar{T}')^2 - 4V(\bar{T})\right] 
       \delta\Phi^2 \\
       && \mbox{} - \left[16 D' (\partial_1\delta\Phi) + 16Q(\partial_2\delta\Phi)
            - 4 \bar{T}' (\partial_1\delta T)\right]\delta\Phi \bigg\} \,.\nonumber
\ee
Defining the canonically normalized fields 
\be
\label{fluctuations}
\phi_1 = 2 e^{-\bar{\Phi}}\delta\Phi \qquad
\phi_2 =e^{-\bar{\Phi}}\delta T\,,
\ee
we can put the Lagrangian density of the fluctuations into the 
form $\phi_i L_{ij} \phi_j$, where
\be
\label{fluctuation_L}
L = 
 \left[
 \begin{array}{cc}
 -\partial^2 + \Delta & 0 \\
 - 2 \bar{T}'(D' + \partial_1) \;\;\; & \partial^2 - \Delta - V''(\bar{T})
 \end{array}
 \right]
\ee
and $\Delta = (D')^2 - D'' + Q^2$.
To find the spectrum of fluctuations we need to diagonalize the above matrix.
This is in general a hard problem, but one eigenmode is easily found:
\be
\label{dilaton_mode}
\phi(x_1,x_2,\dots,x_{D-1}) = \left[
\begin{array}{cc}
e^{-D(x_1)}\tilde{\phi}(x_2,\ldots,x_{D-1}) \\
0
\end{array}
\right] \,.
\ee
This describes a mode of the dilaton which is localized on the soliton,
and which has a mass $Q$. It is tempting to identify it with the dilaton of
the $(D-1)$-dimensional string theory, which has precisely this mass.
However it is not clear whether this mode will survive, and with the same mass,
once the metric fluctuations are added.

\section{Solvable toy models for tachyon solitons}

The system of equations (\ref{gravity})-(\ref{dilaton}) is hard to solve for
a general tachyon potential.
In this section we will consider three exactly solvable toy models
which exhibit the generic features described in the previous section.
In these models we will work in reverse:
we will assume a simple form for the tachyon (or dilaton) profile, and derive the corresponding
tachyon potential and dilaton (or tachyon) profile.

\subsection{Model 1: tachyon kink}

In the first model we assume a kink form for the tachyon
\be
 \bar{T}(x_1) = \beta\tanh{\alpha x_1} \,.
\ee
Note that this would be an exact solution for a decoupled tachyon with a potential
\be
V_0(T) = -\alpha^2 T^2 + {\alpha^2\over 2\beta^2} T^4 \,.
\ee
We expect this to be a good approximation to the potential energy for the above solution
at small $\beta$, since the dilaton scales as $\beta^2$.
We can compute the exact potential as follows.
Dividing the tachyon equation (\ref{tachyon}) by $T'$,
differentiating again, and using (\ref{gravity}) to eliminate $D$ gives
\be
\label{potential_equation}
 \left({T''\over T'}\right)' - \left(T'\right)^2
  - {\partial\over\partial T}\left({V'(T)\over T'}\right) T' = 0 \,.
\ee
Our ansatz for the tachyon satisifies
\be 
 \bar{T}' = {\alpha\over \beta}(\beta^2 - \bar{T}^2) \;,\quad
 \bar{T}'' = {-2\alpha^2\over \beta^2} \bar{T}(\beta^2-\bar{T}^2) \,,
\ee
so we can integrate twice with respect to $T$ to get
\be
 V(T) = {\alpha^2\over 2}\left(-(2+\beta^2)T^2 + {2\beta^2+3\over 3\beta^2}T^4
   - {1\over 9\beta^2}T^6\right) \,.
\ee
The two constants of integration have been determined by the conditions $V'(0)=0$ and $V(0)=0$.
The latter is something we expect generically for closed string tachyons.
We can now solve for the dilaton using either the dilaton
or tachyon equation, to get
\be
D(x_1) = \Phi_0 + {\beta^2\over 3}\left(\ln(\cosh\alpha x_1) -
{1\over 4}\mbox{sech}^2\alpha x_1\right)\,.
\ee
Consistency of the two equations then requires
\be
 Q={\alpha\beta\over 2}\,.
\ee
We see that, as argued on general grounds in the previous section, the dilaton
grows away from the soliton in both directions.
In this case the asymptotic behavior as $x_1\rightarrow \pm\infty$ is linear
\be
D(x_1) \sim {1\over 3}\alpha\beta^2|x_1| \,.
\ee
This should be contrasted with the linear dependence on the soliton worldvolume coordinate
$Qx_2$, which is weakly coupled on one side. Spacetime is effectively localized on 
a semi-infinite part of the $(D-1)$-dimensional soliton.
This is consistent with the identification of the soliton as the linear-dilaton vacuum of the $(D-1)$-dimensional
sub-critical string theory.

\begin{figure}
\begin{center}
$\begin{array}{c@{\hspace{1in}}c}
\epsfig{file=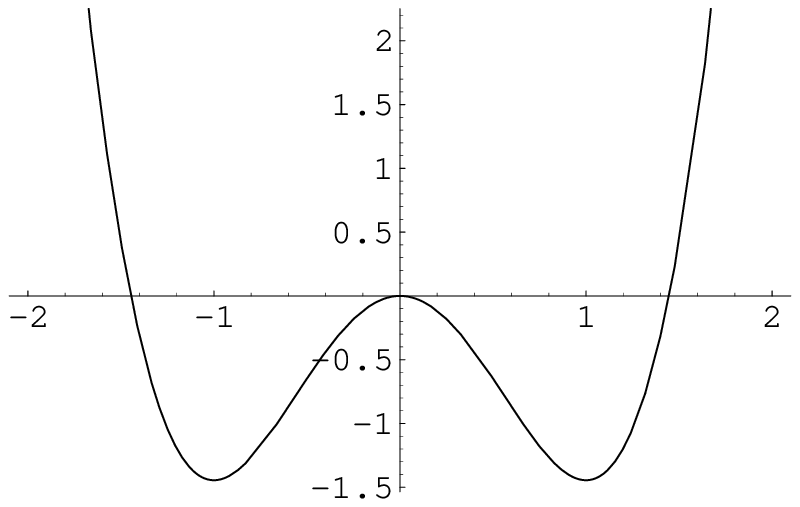, width=2in} &
    \epsfig{file=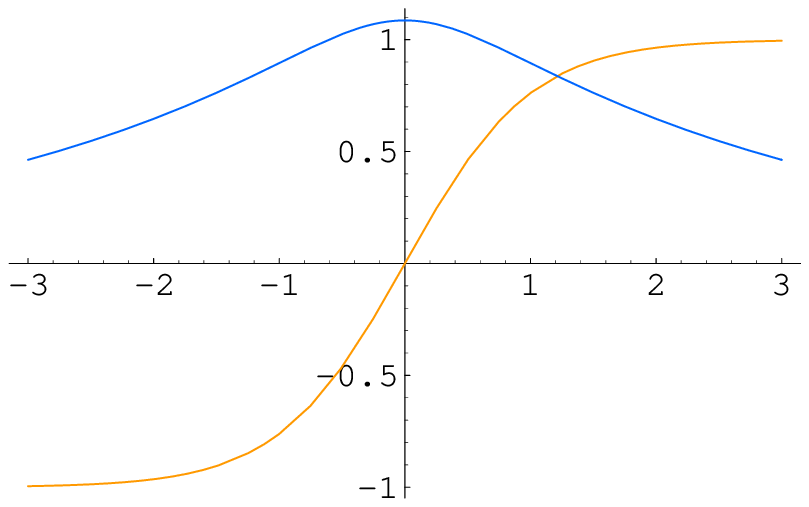,width=2in} \\ [0.4cm]
\mbox{\bf (a)} & \mbox{\bf (b)}
\end{array}$
\end{center}
\caption{{\bf (a)} The potential $V(T)$ with $\beta=1$ for model 1. {\bf (b)} The tachyon profile $\bar T(x_1)$ and the dilaton profile $e^{-D(x_1)}$. }
\label{Toy1}
\end{figure}

\subsection{Model 2: tachyon lump}

In this model we assume a particular lump form for the tachyon
\be
\label{lump}
 \bar{T}(x_1) = -\gamma+\beta\tanh^2\alpha x_1 \,.
\ee
For the special case $\beta=3 \gamma$ this is an exact solution for a decoupled tachyon with 
a potential \cite{Zwiebach:2000dk}
\be
\label{Tcubed}
V_0(T) =  \mbox{} - 2\alpha^2 T^2 
+ {2\alpha^2\over \beta} T^3 \,.
\ee
This should be a good approximation to the potential energy of the lump in the tachyon-dilaton 
system for small $\beta$.
To compute the exact potential we will again use (\ref{potential_equation}).
Defining $\tau\equiv \bar{T}+\gamma$, the lump configuration satisfies
\be
  \bar{T}' = {2\alpha\over\sqrt{\beta}}\sqrt{\tau}\, (\beta - \tau) \;,\quad
  \bar{T}'' = {2\alpha^2\over\beta}(\beta-\tau)(\beta-3\tau) \,.
\ee
Substituting into (\ref{potential_equation}) and integrating twice with respect to $T$
we obtain
\be
\label{lump_potential}
 V(T) &=& \alpha^2 \bigg[2\beta\tau - 4\tau^2 +
   \left({2\over\beta} - {8\beta\over 9}\right) \tau^3 +
   {16\over 15}\tau^4 - {8\over 25\beta}\tau^5 \nonumber \\
 & & \mbox + 2C_1\left({1\over 5}\tau^{5/2} - {\beta\over 3}\tau^{3/2}\right)
  + C_2 \bigg]\,,
\ee
where $C_1$ and $C_2$ are integration constants which are determined
in terms of $\beta$ and $\gamma$ by $V(0)=V'(0)=0$. 
\be
D(x_1) = \Phi_0 + {\alpha\sqrt{\beta}\over 4} C_1x_1 +
  {4\beta^2\over 15}\left(\ln(\cosh\alpha x_1) +
  {3\over 8} \mbox{sech}^4\alpha x_1 - {1\over 4}\mbox{sech}^2\alpha x_1 \right),
\ee
and the consistency of the dilaton and tachyon equations requires
\be
 C_1=0 \;\;,\;\; Q^2 ={1\over 2} \alpha^2C_2 \,.
\ee
The former relates the parameters $\beta$ and $\gamma$ as
\be
\label{betagamma}
 \beta = {{4\over 5}\gamma^3 - 3\gamma\over {4\over 3}\gamma^2 - 1} \,.
\ee
The potential (\ref{lump_potential}) is therefore polynomial, and reduces to the simple
cubic potential (\ref{Tcubed}) for small $\beta$ (and $\gamma$), as expected.
The dilaton is asymptotically given by
$D(x_1)\sim {4\over 15}\alpha\beta^2|x_1|$, so
the region far from the soliton is again strongly coupled.
The picture in the Einstein frame is again that of a spacetime
which is effectively $(D-1)$-dimensional.

\begin{figure}
\begin{center}
$\begin{array}{c@{\hspace{1in}}c}
\epsfig{file=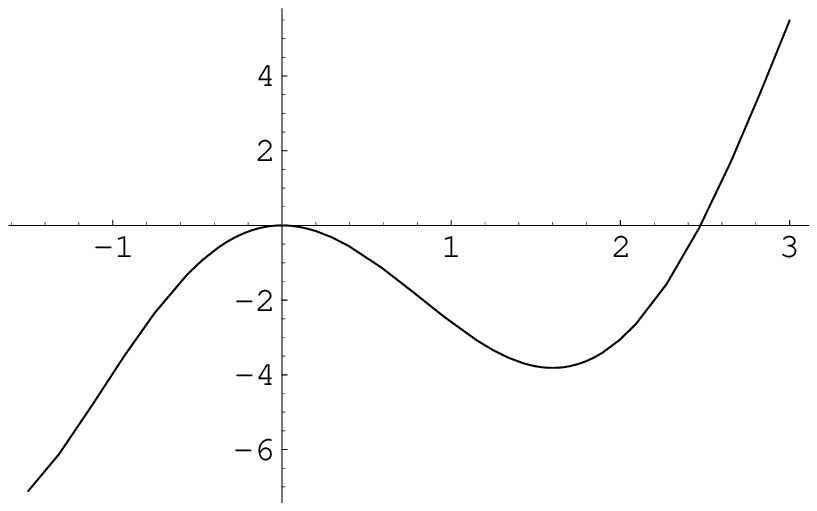, width=2in} &
    \epsfig{file=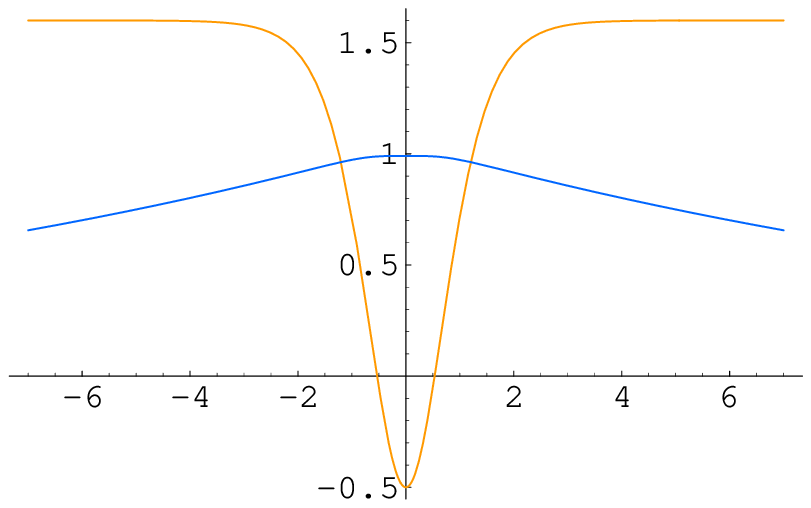,width=2in} \\ [0.4cm]
\mbox{\bf (a)} & \mbox{\bf (b)}
\end{array}$
\end{center}
\caption{{\bf (a)} The potential $V(T)$ with $\gamma=\half$ for model 2. {\bf (b)} The tachyon profile $\bar T(x_1)$ and the dilaton profile $e^{-D(x_1)}$. }
\label{Toy2}
\end{figure}

\subsection{Model 3: dilaton profile}

In this model, rather than assuming a tachyon configuration, we start with a simple dilaton
configuration
\be
\label{dilaton_config}
D(x_1) = \beta\ln(\cosh x_1) \,,
\ee
with $\beta>0$.
Asymptotically the dilaton grows linearly as in the previous examples.
Solving the gravity equation (\ref{gravity}) for the tachyon, with the condition $\bar{T}(0)=0$, gives
a type of kink
\be
\bar{T}(x_1) = \sqrt{8\beta}\arctan\left(\tanh{x_1\over 2}\right) \,.
\ee
The tachyon potenial is found using the dilaton equation (\ref{dilaton}):
\be
\label{potdil1}
V(T)= \mbox{} -\beta(1+2\beta)\biggl(\frac{2\tan(\frac{T}{\sqrt{2\beta}})}
{1+\tan^2(\frac{T}{\sqrt{2\beta}})}\biggr)^2+\beta-2Q^2 \nonumber \,.
\ee
The constants $\beta$ and $Q$ are fixed by the conditions
$V(0)=V'(0)=0$. 
In particular
\be
Q^2 = {\beta\over 2} \,.
\ee
We see again that the slope of the dilaton along the soliton $Q$ is
set by the tachyon potential.

In this model we can also analyze the spectrum of fluctuations
(\ref{fluctuations}) in somewhat more detail, if we ignore 
the tachyon-dilaton mixing term in (\ref{fluctuation_L}).
We will see that the fluctuation spectrum is similar to the one found by Zwiebach
for the open string tachyon soliton in \cite{Zwiebach:2000dk}.
Following Zwiebach's approach,
we expand the fluctuations in terms of modes transverse to the soliton:
\be
\phi_1(x_1,x_2,\dots,x_{D-1}) &=&
\sum \psi_n(x_1)\phi_{1,n}(x_2,\ldots,x_{D-1}) \nonumber\\
\phi_2(x_1,x_2,\dots,x_{D-1}) &=& \sum \chi_n(x_1)\phi_{2,n}(x_2,\ldots,x_{D-1})\,,
\ee
where the modes satisfy the Schrodinger-like eigenvalue equations
\be
\mbox{}-\psi''_n + \Delta\psi_n &=& m_{1,n}^2\psi_n \\
\mbox{}-\chi''_n + (\Delta + V''(\bar{T}))\chi_n &=& m_{2,n}^2\chi_n \,.
\ee
The fields $\phi_{1,n}$ and $\phi_{2,n}$ living on the soliton then
have a mass $m_{1,n}$ and $m_{2,n}$, respectively.
In the dilaton configuration (\ref{dilaton_config}) the eigenvalue equations become
\be
\mbox{}-\psi''_n + \left[\beta^2  - \beta(\beta + 1)\,\mbox{sech}^2 x_1\right] \psi_n &=& 
\left(m_{1,n}^2 - Q^2\right)\psi_n \\
\mbox{}-\chi''_n + \left[\gamma^2 - \gamma(\gamma + 1)\,\mbox{sech}^2 x_1\right]  \chi_n &=& 
\left(m_{2,n}^2 - Q^2 + \delta\right) \chi_n \,,
\ee
where $\gamma(\gamma + 1) = \beta^2 + 5\beta + 2$ and $\delta=\gamma^2 - (\beta + 1)^2$.

The two equations are of the same form as the equation for the fluctuations
of the open string tachyon soliton found in \cite{Zwiebach:2000dk}, and are exactly solvable.
For integer $\beta$ there are exactly $\beta$ discrete bound states for the first equation,
and then a continuum.
More generally the number of bound states is given by the greatest integer less than $\beta$.
Since the ground state $\psi_0$ has a vanishing "energy" eigenvalue \cite{Zwiebach:2000dk}, 
the mass of the lowest lying dilaton mode is $m_{1,0} = Q = \sqrt{\beta/2}$.
This is precisely the mode of (\ref{dilaton_mode}), which we argued is generic.
Surprisingly the mass comes out exactly right for the dilaton field in the linear dilaton
background of the $(D-1)$-dimensional sub-critical string.
This mode will remain unchanged when we turn on the tachyon-dilaton mixing term.

From the ground state of the second equation we similarly find that the lowest lying
mode of the tachyon has a mass squared
\be
m_{2,0}^2 &=& Q^2-\delta \nonumber \\
 &=& \mbox{} - {3\over 2} - {5\over 2}\beta + \sqrt{\beta^2 + 5\beta + {9\over 4}} \,.
 \ee
This mode is tachyonic for all $\beta > 0$, but it is less tachyonic than
the original $D$-dimensional tachyon, which has
$m^2_T = V''(0) = -1-2\beta$. 
This also agrees qualitatively with the interpretation of the soliton as a $(D-1)$-dimensional
sub-critical string theory.
However this mode will mix with the excited dilaton modes
due to the mixing term in (\ref{fluctuation_L}),
so an improved analysis of the coupled fluctuation equations is necessary
in order to determine if its features will persist.

\begin{figure}
\begin{center}
$\begin{array}{c@{\hspace{1in}}c}
\epsfig{file=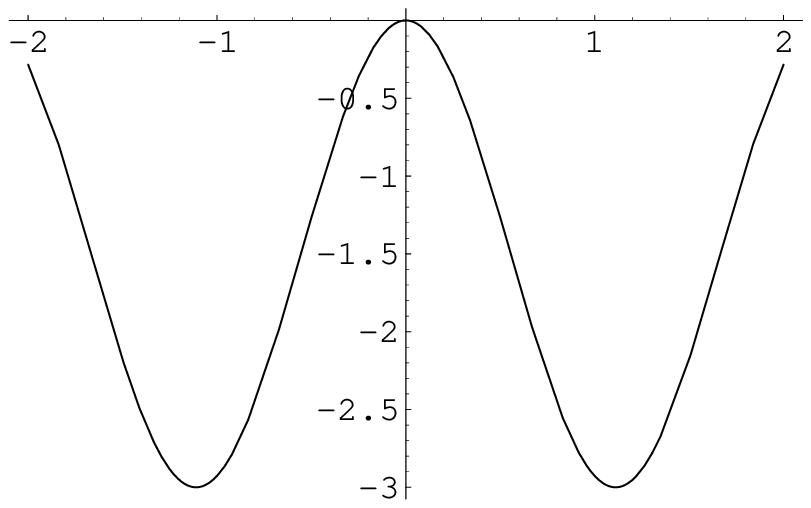, width=2in} &
    \epsfig{file=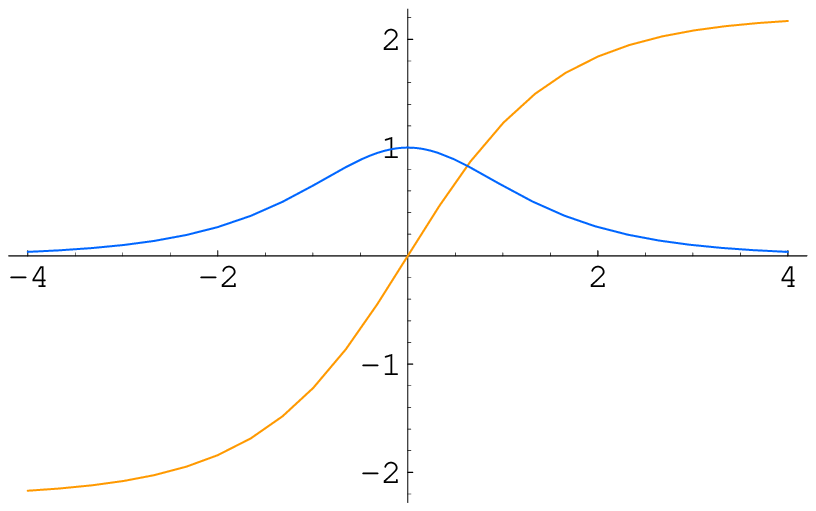,width=2in} \\ [0.4cm]
\mbox{\bf (a)} & \mbox{\bf (b)}
\end{array}$
\end{center}
\caption{{\bf (a)} The potential $V(T)$ with $\beta=1$ for model 3. {\bf (b)} The tachyon profile $\bar T(x_1)$ and the dilaton profile $e^{-D(x_1)}$. }
\label{Toy3}
\end{figure}


\section{Soliton in closed string field theory?}

Ultimately the question of closed string tachyon condensation should be addressed
in the context of closed string field theory (CSFT), where the solitons should correspond
to non-vacuum solutions. These will involve an infinite number of fields.
In particular the potential in CSFT depends on the tachyon, dilaton, and an infinite
set of massive scalar fields. Recently Yang and Zwiebach have made some progress in 
computing this potential in the bosonic CSFT using the method of level-expansion \cite{Yang:2005rx}.
We will use their results to address the question of existence of a co-dimension one soliton
in CSFT.

Other than an infinite number of fileds, CSFT also contains an infinite number of interaction
vertices \cite{Zwiebach:1992ie}. This complicates the level-expansion procedure as compared with 
open string field theory, since one needs to include an infinite number of terms at each level.
Yang and Zwiebach have proposed the following strategy: compute the quadratic and cubic terms
to the highest possible level, and then add quartic terms level by level.
This strategy is supported by the relative weakness of the quartic terms at a given level
compared with cubic terms at the same total level, and suggests that vertices
have an inherent level which grows with the number of legs \cite{Yang:2005ep}.

\subsection{Cubic tachyon potential}  
  
\begin{figure}
\begin{center}
$\begin{array}{c@{\hspace{1in}}c}
\epsfxsize=1.6in \epsffile{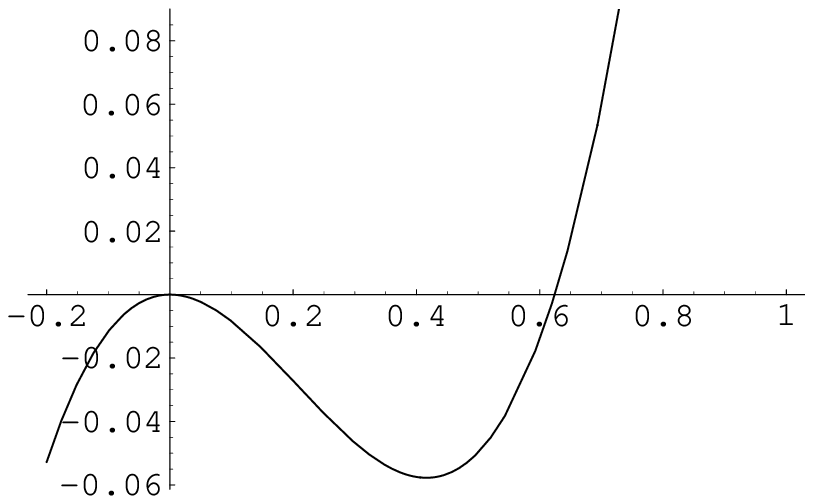} &
    \epsfxsize=1.6in
    \epsffile{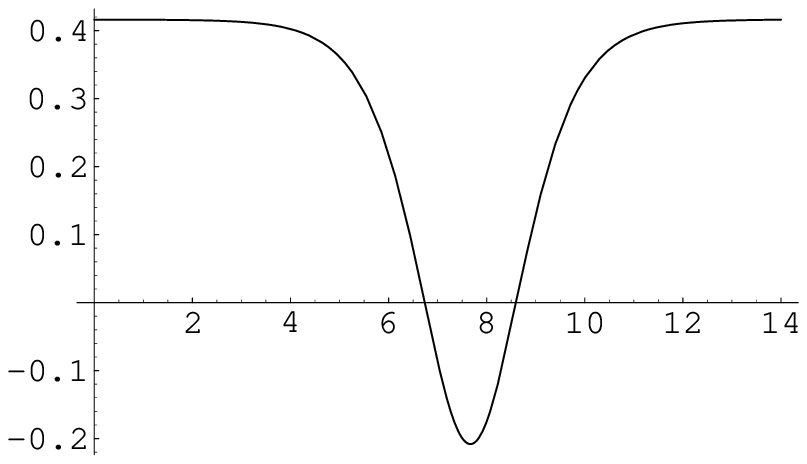} \\ [0.4cm]
\mbox{\bf (a)} & \mbox{\bf (b)}
\end{array}$
\end{center}
\caption{{\bf (a)} The potential $V_0(t)$. {\bf (b)} 
The numerical Lump solution for $V_0$. }
\label{V0pot}
\end{figure}

Adopting the notation of \cite{Yang:2005rx}, the lowest level cubic potential is given by  
($\alpha'=2$)\footnote{We use the lower
case $t$ and $d$ to denote the CSFT tachyon and dilaton, to differentiate them from the 
corresponding sigma model fields.},
\be
\kappa^2 \mathbb{V}_0(t)=-t^2+\frac{3^8}{2^{12}}t^3 \,.
\ee
This has a local minimum at $t_0 = 2^{13}/3^9=0.416$,
where the value of the potential is
\be
\kappa^2 \mathbb{V}_0(t_0) = \mbox{} - {2^{26}\over 3^{19}} = \mbox{} - 0.0577 \,.
\ee
The form of the potential is precisely as in our second toy model (\ref{Tcubed}), 
for which there is an exact lump solution
given by (\ref{lump}) (see Figure~\ref{V0pot}). 
The basic quantity one can compute is the energy density of the soliton.
In the open string tachyon case this quantity was to be compared with the tension 
of the lower-dimensional D-brane. Here one should presumably compare it with
the vacuum energy in the sub-critical linear dilaton background, but one has to
be careful about the precise relation between the different values of $\kappa$ in
the $D$ and $(D-1)$-dimensional theories.
We leave this part for future investigation.
For now we only compute the value of this energy density, or more precisely the difference
between the energy density of the soliton and that of the (true) vacuum,
\be
\Delta{\cal E}_0 = {\cal E}_0(t(x)) - {\cal E}_0(t_0) = 
\int_{-\infty}^{\infty} dx\,\left[{1\over 2}\left(t'(x)\right)^2 + 
V_0(t) - V_0(t_0)\right] \,.
\ee
Note this is energy density per 24-volume, which is actually infinite for both the soliton
and the vacuum, but the difference is finite.
Using translational invariance, and changing the variable of integration to $t$, 
we can rewrite this as
\be
\Delta{\cal E}_0 = 2\int_{t_*}^{t_0} dt\, \sqrt{2\left(V_0(t)-V_0(t_0)\right)}\,,
\ee
where $t_*=-2^{12}/3^9=-0.208$ is the turning point where $V_0(t_*)=V_0(t_0)$.
This can be easily evaluated numerically, giving
\be
\Delta{\cal E}_0 = 0.294 \,.
\ee

\subsection{Yang-Zwiebach potential}

Yang and Zwiebach have computed the cubic potential with fields
up to level 4, and terms up to level 12.
This includes the dilaton $d$ and four massive fields $f_1, f_2, f_3$ and $g_1$,
in addition to the tachyon.
Using the results of Moeller \cite{Moeller:2004yy},
they have also computed all the terms in the quartic potential up to level 4,
as well as the tachyon and dilaton contributions at levels 6 and 8.
Their results are summarized in the appendix.
The soliton will correspond to a non-trivial trajectory in the 6-dimensional field space
which begins and ends at the critical point, and which minimizes the energy density.
To find the exact trajectory we need to solve the coupled system of differential
equations which follow from the YZ potential,\footnote{The kinetic terms of the massive fields
$f_3$ and $g_1$
in the CSFT action have the "wrong" sign, which goes along with their negative mass terms
in (\ref{massive_fields}).}
\be
\partial^2 t - \partial_t V &=& 0 \nonumber\\ 
\partial^2 d - \partial_d V &=&  0 \nonumber\\
\partial^2 f_{1,2} - \partial_{f_{1,2}}V &=& 0 \\
-\partial^2 f_3 - \partial_{f_3}V &=& 0 \nonumber\\
-\partial^2 g_1 - \partial_{g_1}V &=& 0 \nonumber
\ee
Even if we truncate to just the tachyon and dilaton this is a very complicated problem.

\begin{figure}
\begin{center}
$\begin{array}{c@{\hspace{1in}}c}
\epsfxsize=2.3in \epsffile{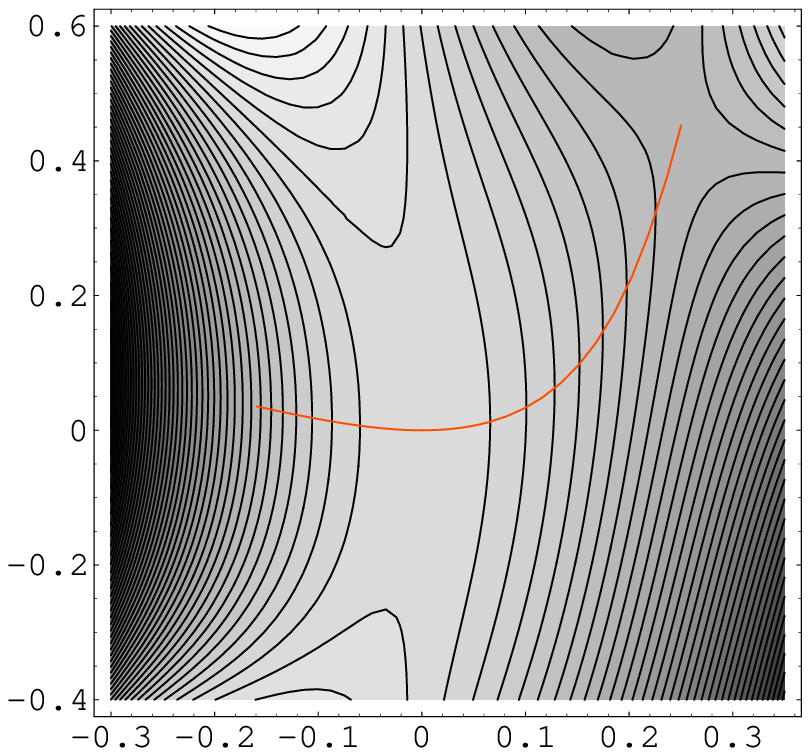} &
    \epsfxsize=2.3in
    \epsffile{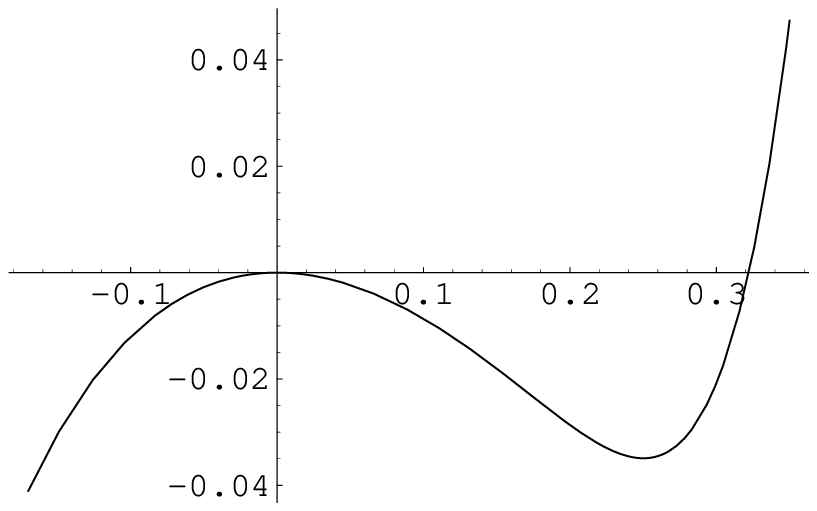} \\ [0.4cm]
\mbox{\bf (a)} & \mbox{\bf (b)}
\end{array}$
\end{center}
\caption{{\bf (a)} Our approximate trajectory is shown on the $(t,d)$ plane
with the effective tachyon-dilaton potential.
 {\bf (b)}  The effective tachyon potential along the trajectory. 
 The minimum is at $t_0=0.250$ and the potential value at the minimum is 
 $V(t_0)=-0.0349$. The turning point is at $t_*=-0.160$. }
\label{V8}
\end{figure}

However, since the tachyon is parameterically small, and 
the other fields all scale as $t^2$,
we can approximate the soliton by solving for the 
dilaton and massive fields in terms of the tachyon using 
\be
\label{trajectory}
\partial_d V\ = \partial_{f_i} V = \partial_{g_1} V = 0\,,
\ee 
and then solving the effective tachyon equation.
This corresponds to a trajectory that starts at the non-trivial critical point,
follows the extremum in the dilaton and massive fields directions
through the critical point at the origin, reaches a turning point, and goes back
the same way.\footnote{We can get an intuitive feel for why this is a good guess by 
thinking about the analog particle
mechanics of a particle moving in the up-side-down effective tachyon-dilaton potential 
$V(t,d)$ obtained by solving for just the massive fields. The trajectory defined by $\partial_d V=0$
follows the dilaton valley from the top of the tachyon hill, through the bottom, to another top,
and back.}
Taking the quadratic and cubic terms to level eight,
and the quartic terms to level four, we can solve numerically for the trajectory
and the corresponding tachyon effective potential, shown in Figure~\ref{V8}.
A numerical computation of the energy density of the lump then yields
\be
\Delta{\cal E} =0.160 \,.
\ee

\section{Discussion}

We have presented evidence that closed string tachyon solitons,
if they exist, describe lower-dimensional sub-critical closed string backgrounds,
similar to the way open string tachyon solitons describe lower-dimensional D-branes.
Configurations corresponding to tachyon lumps or kinks in
the low-energy effective gravity-dilaton-tachyon theory generically have
a flat string-frame metric, and a dilaton which grows in both directions
away from the core. Spacetime is therefore effectively localized on the
$(D-1)$-dimensional worldvolume of the soliton.
In addition, the dilaton depends linearly on the worldvolume coordinates,
which agrees with its expected behavior in the sub-critical theory.
One needs to understand better the spectrum of fluctuations
around the solitons to decide whether it agrees, even qualitatively, with
that of the sub-critical string. Ignoring the gravity fluctuations, we find
a particular mode of the dilaton which has the correct mass to be the 
dilaton in the sub-critical string.

The existence of closed string tachyon solitons depends on the precise form of
the tachyon potential. We gave three toy models for tachyon potentials which
admit a co-dimension one tachyon soliton, and showed that they satisfy
the general properties above.
However to determine the actual tachyon potential of the closed bosonic string
we must turn to closed string field theory.
The lowest-level cubic tachyon potential admits a simple lump solution.
Using some recent advances in the computation of the potential 
to higher levels and vertices, we have argued that the lump solution persists,
although we have not found the exact solution.
Further progress may be possible using multi-field soliton techniques.

By analogy with open string tachyons,
the closed string tachyon conjecture suggests that the solitons
should have an energy density given by the vacuum energy
density, {\em i.e.} cosmological constant, of the linear-dilaton sub-critical
string background,
\be
\Lambda_D = {1\over 2\kappa_{D-26}^2} {D-26\over 3\alpha'} \,.
\ee
In fact the first part of the conjecture would naively
imply that the energy density of the vacuum is given 
by $\Lambda_0$.
What we need to compare for a co-dimension $k$ soliton is then
\be
\Delta{\cal E}_k \stackrel{?}{=} \Lambda_{26-k} - \Lambda_0 \,.
\ee
A naive comparison for both the vacuum and the lump shows they are far off.
However the left hand side involves the 26-dimensional gravitational coupling
of the critical string $\kappa$, whereas the right hand side involves lower-dimensional
gravitational couplings. We need to take this into account before we make the 
comparison. We leave this for future investigation.


\section*{Acknowledgments}
This work was supported in part by the
Israel Science Foundation under grant no.~568/05.

\appendix
\section{The Yang-Zwiebach potential}

The cubic potential to level 12, with fields to level 4, is given by \cite{Yang:2005rx}:
\be
\mathbb{V}^{(3)}_{12} = V_0^{(2)} + V_0^{(3)} + V_4^{(3)} + V_6^{(3)} + V_8^{(2)}
+ V_8^{(3)} + V_{10}^{(3)} + V_{12}^{(3)}
\ee 
where the different terms correspond to the contributions 
to the vertex (quadratic or cubic) given by the superscript, 
with a total level given by the subscript. These are:
\be
\kappa^2V_0^{(2)} = -t^2 \;\; , \;\; \kappa^2 V_0^{(3)} = {3^8\over 2^{12}} t^3
\ee
\be
\kappa^2 V_4^{(3)} =  -{27\over 32}d^2 t + 
\left( {3267\over 4096}f_1 + {114075\over 4096}f_2 + {19305\over 2048}f_3\right) t^2 
\ee
\be
\kappa^2 V_6^{(3)} = -{25\over 8}g_1 t d 
\ee
\be
\label{massive_fields}
\kappa^2 V_8^{(2)} &=& f_1^2 + 169 f_2^2 - 26 f_3^2 - 2 g_1^2 
\ee
\be
\kappa^2 V_8^{(3)} &=& -{1\over 96} f_1 d^2 - {4225\over 864} f_2d^2
+ {65\over 144} f_3d^2 \nonumber \\[5pt]
 &+& {361\over 12288}f_1^2t + {511225\over 55296}f_1f_2t + {57047809\over 110592}f_2^2t
 + {470873\over 27648}f_3^2t -{49\over 24}g_1^2t \nonumber\\[5pt]
 &-& {13585\over 9216}f_1f_3t - {5400395\over 27648}f_2f_3t 
\ee
\be
\kappa^2 V_{10}^{(3)} &=& -{25\over 5832}\left(361f_1 + 4225f_2 - 2470f_3\right)dg_1
\ee
\be
\kappa^2 V_{12}^{(3)} &=& {1\over 4096}f_1^3 + {74181603769\over 26873856}f_2^3
-{31167227\over 3359232}f_3^3\nonumber\\[5pt]
& + & {1525225\over 8957952}f_1^2f_2 -{1235\over 55296}f_1^2f_3 
+ {6902784889\over 80621568}f_2^2f_1 \nonumber\\[5pt]
&-& {22628735129\over 13436928}f_2^2f_3 + {1884233\over 2239488}f_3^2 f_1
+ {4965049817\over 20155392}f_3^2f_2 \nonumber\\[5pt]
&-& {102607505\over 6718464}f_1f_2f_3
-{961\over 157464}f_1g_1^2 - {207025\over 17496}f_2g_1^2 +{14105\over 26244}f_3g_1^2
\ee
The quartic vertices have been computed to level 4 completely, and at levels 6 and 8
for the tachyon and dilaton only,
\be
\kappa^2 V_0^{(4)} &=& -3.1072 t^4 \\
\kappa^2 V_2^{(4)} &=& 3.8721 t^3 d \\
\kappa^2 V_4^{(4)} &=& 1.3682t^2d^2 + t^3\left(-0.4377f_1-56.262f_2+13.024f_3
+0.2725g_1\right) \\
\kappa^2 V_6^{(4)} &=& -0.9528 td^3  + \cdots\\
\kappa^2 V_8^{(4)} &=& -0.1056 d^4 + \cdots 
\ee

\end{document}